\documentclass[12pt]{article}
\usepackage{graphicx}  
\usepackage{graphics}  
\usepackage{amssymb}
\usepackage{amsmath}
\usepackage{epsfig}
\begin{document}

\title{Spin Dependent Thermoelectric Currents of Tunnel Junctions, Small Rings and Quantum Dots: Onsager Theory}

\author{K.H.Bennemann,\\ Institute for Theoretical Physics, Freie Universit\"{a}t Berlin,\\ Arnimallee 14, 14195 Berlin}
\maketitle

\abstract{Spin Currents in Tunnel Junctions for example induced by thermoelectric forces due to temperature and magnetization gradients etc. are analyzed. Using Onsager theory in particular for magnetic tunnel junctions, metallic rings and quantum dots yields directly,   spin dependently all thermoelectric and thermomagnetic effects like Seebeck and Peltier ones and Josephson like Spin currents driven by the phase gradient of the magnetization. The results can be compared with recent experiments determining the Spin dependent Seebeck effect and other thermoelectric effects. The Onsager theory can be extended towards an electronic theory by expressing the Onsager coefficients by current correlation functions and then calculating these using Lagrange formalism, symmetry and scaling analysis. Note, Onsager theory can also be applied to spin currents in molecules and in magnetic ionic liquids.}
\newpage
\tableofcontents

\section{Introduction}
Recently, spin dependent currents in nanostructures and tunnel junctions
have been discussed intensively \cite{1,2,3,4,5}. In particular
the spin dependent thermoelectric and thermomagnetic effects like Seebeck effect and the
heat due to spin dependent currents in ferromagnets, spin Peltier effect
receive special attention \cite{4,5}. The interdependence of the various currents is most interesting and well described by Onsager theory. Onsager theory $j = L X$ for the currents j driven by the spin dependent generalized thermodynamical forces X like temperature gradient or magnetization gradient etc. yields the spin dependent thermoelectric effects \cite{6}. In particular this holds for nanostructures like tunnel junctions and metallic rings \cite{2,7} and tunnel currents through molecules and spin currents in magnetic ionic liquids. Note,
even if originally one has a homogeneous magnetization a temperature gradient $\Delta T$ will induce a difference $\Delta M$ in the magnetization
and $\Delta M \propto \Delta T$.
\begin{figure}
\centerline{\includegraphics[width=.5\textwidth]{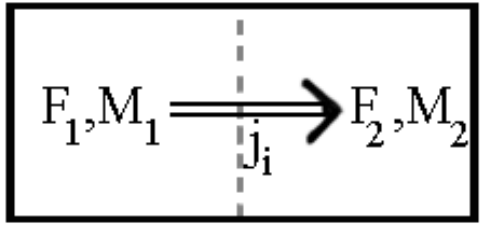}}
\caption{Illustration of an inhomogeneous ferromagnetic tunnel junction with temperature gradient $\Delta T=T_1-T_2$ and magnetization gradient $\Delta M(t)= M_1 - M_2$ and possibly other gradients. Coupled spin dependent currents $j_i$ are expected in accordance with Onsager theory. Note the currents may induce a temperature gradient and thus affect the gradient $\Delta M(t)$ between ferromagnets $F_1$ and $F_2$. In particular one gets Josephson like spin currents due to a phase difference of the magnetizations ($j_s^J\propto\frac{dM}{dt}\propto\protect\overrightarrow{M_1}\times\protect\overrightarrow{H_{eff.}}+...$) and for example from the coupled currents a spin dependent Seebeck effect ($\Delta (\mu_\uparrow - \mu_\downarrow) \propto \Delta T$ and Peltier effect ($heat \propto$ current). Clearly the various currents will depend on the magnetic configuration of $M_1$ and $M_2$, ferromagnetic vs. antiferromagnetic configuration of the two magnets and thus also the electric potential gradient $\Delta \varphi$ depends on $\protect\overrightarrow{M_1}$ and $\protect\overrightarrow{M_2}$), ($\Delta \varphi \propto \Delta T$).}
\end{figure}

Also, of course, even in a homogeneous ferromagnet one gets for itinerant electrons $j_\uparrow \neq j_\downarrow$ for the spin currents due to the spin dependent density of states ($N_\uparrow (\varepsilon)\neq N_\downarrow(\varepsilon)$) etc.. Already the Boltzmann equation yields this in a qualitative correct way.

In analogy to the Josephson current
in superconductors due to the phase difference of the order parameter,
one expects also for ferromagnets, magnetic tunnel junctions, with a
gradient in the magnetization its magnitude and phase ($M=|M|\exp (i\phi)$)
a similar Josephson like spin current \cite{2,3,6}. Such currents are also expected for metallic rings with inhomogeneous magnetization.

Of course, in
inhomogeneous ferromagnets, see Fig.1, coupled currents involving spin
and charge are expected. This is elegantly described by Onsager theory,
see also Bennemann \cite{3}. Note, the magnetization may result from local
magnetic moments (for example, in rare--earth) or from spins of itinerant electrons
in transition metals or rare--earth.

The Onsager theory may also be applied to describe thermoelectric and thermomagnetic effects in magnetic ionic liquids. Then spin dependent pressure effects are expected in case of pressure gradients, possibly interfering with other gradients.

Interesting are in particular inhomogeneous systems (nanostructures, tunnel junctions) like ($FM_1|N|FM_2$), ($FM|SC|FM$), etc. for studying spin lifetimes of electron spins injected from a ferromagnet (FM) into a nonmagnetic metal (N) and for studying spin currents in superconductors (SC), analysis of singulet vs. triplet superconductivity, (FM|SC) interfaces \cite{2,7}. This may be used to test triplet superconductivity.

As indicated already by the giant Faraday effect in graphene and a few layers of graphene one expects for graphene structures due to the relatively long spin mean free paths interesting spin dependent thermoelectric effects (see MacDonald {\it et al.}, Bennemann and others). For example, for tunnel junctions involving graphene between the two ferromagnets the Josephson like spin current driven by a gradient of the phase of the magnetizations on the left and right side of the tunnel junction could be observed.

For tunneling involving triplet superconductivity and ferromagnetism the interplay of the order parameters yields novel properties of tunnel junctions. For example, one gets Cooper pair tunneling even for no phase difference between the superconductors on both sides of the tunnel junction
\cite{7}. Note, regarding the Josephson like spin current driven by the phase gradient of the magnetization on the left side and right side of
the tunnel junction \cite{2,6}, see Figure 2, this might require relatively long spin mean free paths. Thus, weak spin--orbit scattering and tunneling for example through graphene favors this spin current. Strong spin--orbit scattering is expected to suppress this Josephson spin current.

Spin currents in metallic rings, in particular persistent ones, are interesting. One expects that the Aharonov--Bohm effect, spin--orbit coupling and interferences of magnetism and superconductivity yield novel behavior \cite{8}.

For fluctuating spin currents (in z--direction) one gets according to the Maxwell equations also accompanying electromagnetic fields, see $4\pi j_{s,z}=-4\pi\mu_B\partial_t\langle S_z\rangle = \partial_x E_y-\partial_y E_x$, etc. Generally the connection between spin currents and magnetization dynamics is given by $\partial_t M_i + \partial_\mu j_{i\mu,\sigma} = 0$ \cite{2}. According to Kirchhoff for example the emissivity (e) of a tunnel junction (or thin film) is related to the magnetization dynamics and magnetic resistance ($\Delta e/e \simeq a (GMR)$, where $\Delta e$ is the change in the emissivity due to changing the magnetic configuration $(\uparrow/\uparrow)$ to $(\uparrow/\downarrow)$ of neighboring thin films or tunnel junctions, GMR
is the giant magnetoresistance).

In general nonequilibrium thermodynamics describes the thermoelectric and thermomagnetic effects. The currents $j_i$, including spin currents,
are driven by the spin dependent thermoelectric forces $X_i = - \frac{\partial\Delta S}{\partial x_i}$, where S is the entropy and $\dot{x_i} = j_i$ ($x_i$ = fluctuations of usual thermodynamical variables, $x_i$ and $X_i$ are conjugate variables). Thus, the currents can be calculated from
\begin{equation}
    j_i \propto (1/X_i) d \Delta F / d t  \quad ,
\end{equation}
where F is the free--energy determined for example by an electronic hamiltonian. Note, Eq.(1) is of basic significance, since it relates the currents to the free--energy, see F.Bloch, S.de Groot and others \cite{8,9}. Hence, the currents may be calculated from the free--energy and this permits obviously application of scaling theory regarding phase transition behavior.

In case of itinerant electrons the spin dependent currents result from the gradients $\Delta \mu_\sigma$ of the spin $\sigma$ dependent chemical potentials $\mu_\sigma$.
Note, $\mu_\uparrow - \mu_\downarrow \simeq 2 \mu_0 H_{eff}$, where $H_{eff}$ is the effective molecular field acting on the itinerant spins \cite{9}, $\mu_{\uparrow(\downarrow)}=\mu\mp \mu_0 H_{eff}$.

Above Eq.(1) follows from $d\Delta S=\Sigma_i X_ix_i$, $j_i = \dot{x_i}$ and then
\begin{equation}
     d\Delta \dot{F} = - \sum_i (T j_i X_i) + ... \quad.
\end{equation}
This yields in particular $j_i \Delta \varphi = - d\dot{F}$ \cite{8}. As discussed later and as quantum mechanically expected of course the phase of the driving force
$X_{i}(t)$ plays an important general role, see for example Josephson currents in superconductors or spin currents in magnets, etc..
Clearly, in general Eq.(1) includes also contributions due to time dependencies of phases occurring in the free--energy and applies also to superconductors.

The following study may be useful to demonstrate how Onsager theory yields the interdependence of the various currents ( in nanostructures ). Onsager theory is most useful to describe directly all thermoelectric effects etc. , spin dependently. Already known results \cite{1,4,5} and new results \cite{2} are presented. This may help to apply studies by F.Bloch \cite{8} and others to spintronics and to new problems.

\section{Theory}

\subsection{Onsager Theory}
As a general framework for deriving the coupled spin dependent currents in tunnel junctions (and nanostructures in general) driven
by thermoelectric forces $X_i$ like magnetization--, temperature-- or
chemical potential--gradients one may use Onsager theory, see Kubo,
Landau, de Groot {\it et al.} \cite{9}.

Generally for deriving the spin dependent thermoelectric and thermomagnetic currents the Onsager theory is most direct and useful. Then the coupled spin dependent currents $j_i$ are given according to Onsager theory by (expanding $\dot{x_i}=f({x_l})$)
\begin{equation}
       j_i(t) = L_{ij} X_j(t) + L_{ijl} X_j X_l +...,
\end{equation}
with driving forces \cite{9,10} $X_i=-\partial\Delta S/ \partial x_i$ and using for the entropy S the expression $d\Delta S=\Sigma_i X_ix_i.$ Note, $x_i$ denotes the extensive thermodynamical variables like $E,$ $V,$ $e$ etc. Then
from thermodynamics one gets
\begin{eqnarray}
d\Delta S = \Delta(1/T)dE &+& \Delta(p/T)dV - \Sigma_{\sigma}\Delta(\mu_{\sigma}/T)dN_{\sigma}\nonumber \\
            &+& \Delta(H^{'}_{eff}/T)dM_L,
\end{eqnarray}
where $\mu_{\sigma}$ is the spin dependent chemical potential of itinerant electrons, $M_L$ the magnetization of local magnetic moments and $H^{'}_{eff}$
the effective molecular field acting on the local spins. The spin $\sigma$ dependent chemical potential is given by
\begin{equation}
     \mu_\sigma = - e \varphi + \mu(0)-\sigma \mu_0 H_{eff},
\end{equation}
where $\varphi$ is the potential acting on the electron charge e and where $H_{eff}$ (including an external magnetic field H) is the molecular field acting on the itinerant electron spins with magnetization M and $\mu(0)$ the chemical potential for $H_{eff}=0$, $H_{eff} = H + q M $, respectively. Note, the term $\Delta\mu_{\sigma}dN_{\sigma}$ can also be put into the form $(\Delta (\mu(0)-e\varphi) dN - \sigma \mu_0 \Delta H_{eff})$. The spin polarized electrons or ions in gases and liquids may cause a partial pressure $p_\sigma$ and this could be included in the Onsager Eqs.. We may put $X_1\equiv X_E = \Delta T /T^2$, $X_{2\sigma} \equiv X_\sigma=\Delta(\mu_{\sigma}/T)$,
$X_3 = X_M = - \Delta(H^{'}_{eff}/T)$, $X_{4\sigma}= - \Delta (p_\sigma/T)$, the partial pressure of the electrons with spin $\sigma$  (in liquids, magnetohydrodynamics etc. ), $X_5 = - \Delta (p/T)$, etc.

Thus, one finds for the coupled currents $j_i=L_{ij}X_j+...$ driven by the forces $X_i$ with $i=1=E$ referring to the energy (heat) current,  i=2=$\uparrow$ spin current, i=3=e (electric current), etc.
the expressions \cite{9}
\begin{eqnarray}
      j_E &=& L_{11} \Delta T/T^{2}  + \Sigma L_{12}^\sigma \Delta(\mu_{\sigma}/T) - L_{13} \Delta(H^{'}_{eff}/T)\nonumber \\ &-&
            L_{14}^\uparrow \Delta (p_\uparrow/T)
             + L_{14}^\downarrow \Delta (p_\downarrow/T) + ...,
\end{eqnarray}
\begin{eqnarray}
      j_\uparrow &=& L_{21} \Delta T/T^2  + L_{22}^\uparrow \Delta(\mu_{\uparrow}/T) - L_{23} \Delta(H^{'}_{eff}/T)\nonumber \\ &-&
                    L_{24}^\uparrow\Delta (p_\uparrow/T) + ...,
\end{eqnarray}
\begin{eqnarray}
      j_e = j_\uparrow + j_\downarrow = (L_{21}^\uparrow + L_{21}^\downarrow) \frac{\Delta T}{T^2} + ... ,
\end{eqnarray}
\begin{equation}
       j_s = j_\uparrow - j_\downarrow = (L_{21}^\uparrow - L_{21}^\downarrow) \frac{\Delta T}{T^2} + ... , (spin - current),
\end{equation}
and for the local moment magnetization the current
\begin{equation}
      j_{M_L} = L_{31} \Delta T/T^2 + \Sigma L_{32}^\sigma \Delta(\mu_{\sigma}/T) - L_{33} \Delta(H^{'}_{eff}/T)+... \quad.
\end{equation}
Note, the replacement $\uparrow \rightarrow \downarrow$ yields $j_\downarrow$. The spin currents $j_\uparrow$ and $j_\downarrow$ may be coupled by spin flip processes, in particular spin--orbit interaction. Then a term proportional to
$\Delta \mu_\downarrow$ could also contribute to
$j_\uparrow$. As usual symmetries may reduce the number of different Onsager coefficients $L_{ij}$, for example  $L_{ij}(H)=L_{ji}(-H)$ may hold etc.. The Onsager coefficients may be expressed by the experimentally observed transport
coefficients \cite{9}.

The most important and central property of the Onsager equations is the interdependence of the vaious currents driven by the forces $X_i$. In particular the driving force
\begin{equation}
        X_{2\sigma} = \Delta (\mu_\sigma / T) \propto - \Delta (H_{eff}/T) + ...  \propto - \Delta (M/T) + ...
\end{equation}
causes correlated currents due to gradients of the magnetization with respect to phase and magnetization magnitude, respectively
($\Delta (\mu_\sigma /T)= \frac{1}{T}\Delta _\sigma (T) - \mu_\sigma \Delta T/T^2 $). The phase gradient driven spin currents are of Josephson type \cite{2,6}. The Onsager equations show that the spin Josephson current is accompanied by a contribution to $j_e$, $j_E$, for example, or better $\Delta M$ due to a phase gradient induces also a contribution to
the other currents, $j_e$ etc.. This is immeadiately obvious from Onsager theory and yields new behavior.

Note, the Onsager Eqs. apply also to superconductors (use for example the two fluid model for superfluids) and yield different behavior for the single particle currents regarding singulet
and triplett superconductors, in particular for $j_e$ and $j_\uparrow$. The current of the Cooper pairs may be added to above Onsager equations. In case of triplett pairing the spin or angular momentum current of the Cooper pairs is of particular interest.

Also note, the Onsager theory applies to ions (in liquids, gases, see magnetohydrodynamics), in particular magnetic ones. A special interesting application
of Onsager theory may be to a lattice of atoms or molecules and of quantum dots and possibly magnetic currents in intergalactic space.

The Onsager equations are very useful for deriving directly the thermoelectric and thermomagnetic effects. The coefficients in the Eqs. need be determined experimentally, by various conductivities, and may be calculated from the free--energy F and using for F an electronic theory. Then scaling theory may be applied to the coupled currents near phase transitions.

Special situations are easily described by the Onsager equations. For example, decoupling of charge and spin current is described by \begin{equation}
      j_e = 0, j_\sigma \neq 0, j_s \neq 0.
\end{equation}

Taking into account the spatial anisotropy induced by the molecular field $\overrightarrow{H_{eff}}$ and by an external magnetic field H one has $j_{i}^{\alpha}$, $\alpha=x,y,z$, denoting the current of
sort i in the direction $\alpha$. The situation simplifies for $\overrightarrow{H_{eff}} \bot (x,y)$ and isotropic plane (x,y) implying symmetries
for coefficients $L_{ij}$ upon transformation $x\protect\rightleftarrows y$, see for example de Groot {\it et al.} \cite{9}. For spatial anisotropy due to H and $\overrightarrow{H_{eff}}$ one has ($j^\alpha=\Sigma_\beta (L^{\beta})X^\beta+...$, $\alpha,\beta=x,y$, $(L^\beta)$ is the coefficient matrix)
\begin{equation}
         j_{i}^{\alpha} = L_{ij}^{x} X_{j}^x + L_{ij}^{y} X_{j}^y,     \alpha = x, y.
\end{equation}
Symmetry with respect to $x \rightleftharpoons y$, $H \rightarrow -H$, etc. will reduce the number of different Onsager coefficients as usual. Note, the coupling of responses in x-- and y--direction. It is $X_{E}^{\alpha}=-1/T^2 \Delta_\alpha T$,
$X_{\sigma}^x=\Delta_{x}(\mu_{\sigma}/ T)$, etc. . This yields then spin dependent galvanomagnetic effects, Hall--effect ($\Delta_{y} \mu$ due to currents $j_{i}^x$, etc.), isothermal Nernst effect
($\Delta_{y} \mu$ due to energy current $j_{E}^x$), etc.\cite{9}.

Thus, for example
in presence of the external magnetic field H (or molecular field $H_{eff}$) perpendicular to the currents in a tunnel junction and taking into account the induced anisotropy, one gets from
above Onsager equation $j_1\equiv j_E, j_2\equiv j_e$  $(j_E = j_{E\uparrow} + j_{E\downarrow}), \Pi_\sigma = j_{E\sigma}/ j_\sigma)$
\begin{equation}
j_E^x = \Pi_{\uparrow}^x j_{\uparrow}^x + \Pi_{\downarrow}^{x} j_{\downarrow}^x + ...
\quad ,
\end{equation}
and
\begin{eqnarray}
j_e^x &=& L_{21}^x \frac{\Delta_x T}{T^2} + L_{21}^y \frac{\Delta_y T}{T^2} \nonumber \\ &+& L_{22}^{x\uparrow} \Delta_x (\mu_{\uparrow} /T) +
        L_{22}^{x\downarrow}\Delta_x(\mu_{\downarrow} /T +         L_{22}^{y\uparrow} \Delta_y(\mu_\uparrow/T) + ....\quad,
\end{eqnarray}
and similar Eqs. for $j_e^y$, for spin currents $j_{\sigma}^x$, $j_{\sigma}^y$, etc.. Note, x$\rightarrow$y yields $j_E^y$, etc.,
see de Groot and others \cite{9}.

One gets from these Eqs. as expected that currents induce a spin voltage (Hall--effect) etc.. For example, the spin current $j_s^x=j_{\uparrow}^x-j_{\downarrow}^x$
in x--direction induces the spin voltage
\begin{equation}
            \Delta_y (\mu_\uparrow - \mu_\downarrow) \propto j_s^x ,   \Delta_y \mu_\sigma \propto j_\sigma^x
\end{equation}
in y--direction.

Some magnetogalvanic effects are discussed later. First spatial anisotropy due to
the field $\overrightarrow{H_{eff}}$ is not explicitly taken into account.

Spin Currents $j_i \sim \Delta M(t,..)$ :

The spin currents in tunnel junctions resulting from $\Delta M(t)$ with respect to phase gradient and gradient of the magnitude of
the magnetization are of special interest. The Onsager Eqs. indicate immeadiately that $\Delta M$ affects the various currents.
As mentioned already the spin currents driven by the phase gradient of the magnetization of the itinerant electron spins and of the local magnetic moments may also be derived from the continuity equation for the magnetization and from the Landau--Lifshitz equation \cite{2}. In particular as discussed already the gradient of the magnetization phase yields the Josephson like spin currents
between two ferromagnets 1 and 2, in tunnel junctions and at interfaces, see Fig.1 and previous Eqs. \cite{11}.
\begin{figure}
\centerline{\includegraphics[width=.5\textwidth]{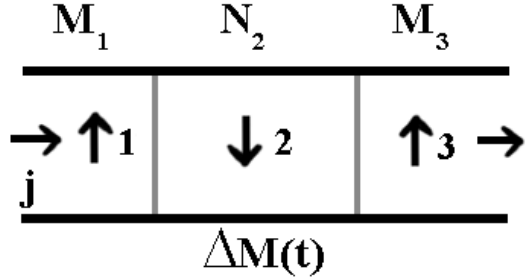}}
\caption{Nanostructure, tunnel junction composed of ferromagnetic metals 1 and 2. Due to the magnetization phase difference $\Delta \phi$ between
ferromagnet 1 and 3 a spin current proportional to $\sin \Delta \phi$ is expected. Spin relaxation controls this current. Note, if a ferromagnetic metal 2 is put in between metal 1 and 3 with a.f. oriented magnetization causing a giant magnetoresistance then one may generate an ultrafast oscillating (modulated) current by photon irradiation changing the magnetization in part 2 by $\Delta M(t)$ time dependently. According to Onsager theory the $\Delta M$--gradients are the driving forces of the spin currents and interference effects may occur. A Josephson like spin current due to the gradient $\Delta \phi$ results also if $N_2$ is non--magnetic and if the spin mean free path is comparable or longer than the thickness of $N_2$. ($N_2$: graphene). Magnetization dynamics may cause ultrafast oscillating radiation according to Maxwell
equations. As indicated by the giant magnetoresistance (GMR) or tunnel resistance (TMR) the tunnel currents depend on the relative orientation of the magnetizations. The Seebeck and Peltier effect will reflect this.}
\end{figure}

Regarding the response to a gradient in the phase of the magnetization, one gets from the gradient of the phase of the magnetization for a tunnel junction or for film multilayers a spin current. Using the continuity equation $\partial_t M_i + \partial_\mu j_{i\mu,\sigma} = 0$ under certain conditions
or using the Landau--Lifshitz equation $dM/dt=a \overrightarrow{M}\times\overrightarrow{H_{eff}}+...$, where $\overrightarrow{H_{eff}}$ refers to the effective molecular field, one may derive a spin current including a Josephson like spin current $j^J$ of the form
\begin{equation}
     j_\sigma=j^1_{\sigma}(\varphi)+j^J.
\end{equation}
Here, $j^1_{\sigma}(\varphi)$ is the spin current due to the electrical potential $\varphi$ and may for example result from the spin dependent density of states. The Josephson like spin current driven by a phase gradient $\Delta \phi$ of the magnetization is given by
\begin{equation}
       j^J\propto dM/dt \propto \overrightarrow{M_L}\times\overrightarrow{M_R}+...\propto |M_L||M_R|\sin(\phi_L - \phi_R +...),
\end{equation}
where $(\phi_L-\phi_R)$ is the phase difference of the magnetization on the left and right side of a tunnel junction (or of two films). Note, damping of spin transport may approximately be taken into account, see Landau--Lifshitz equation or Landau--Lifshitz--Gilbert equation, in the coefficient in front of the term ($\overrightarrow{M_L}\times \overrightarrow{M_R}$).

For later discussion one writes $\Delta (\mu_\sigma/ T) = (\Delta\mu_\sigma) /T - \frac{\mu_\sigma}{T^2} \Delta T$ and then the Onsager equations for the coupled currents of the itinerant electrons may be rewritten as
\begin{eqnarray}
j_E &=& (L_{11}- \Sigma_\sigma L_{12}^\sigma \mu_\sigma)\frac{\Delta T}{T^2} + \Sigma_\sigma L_{12}^\sigma \Delta\mu_\sigma / T
            + ... \nonumber \\   &=& (j_E / j_e) j_e    = \Pi_\uparrow j_\uparrow + \Pi_\downarrow j_\downarrow , \nonumber \\
j_e &=& C_e \frac{\Delta T}{T^2} + \frac{1}{T}[L_{22}^\uparrow \Delta(\mu_\uparrow + \mu_\downarrow) + (L_{22}^\downarrow - L_{22}^\uparrow) \Delta \mu_\downarrow] +... ,\nonumber \\
j_s &=& C_s \frac{\Delta T}{T^2}
+ (1/T) [ L_{22}^\uparrow \Delta (\mu_\uparrow - \mu_\downarrow) - (L_{22}^{\downarrow} - L_{22}^{\uparrow}) \Delta \mu_\downarrow ] ,
\end{eqnarray}
with coefficients
\begin{eqnarray}
    C_e &=& [(L_{21}^\uparrow + L_{21}^\downarrow) - (L_{22}^\uparrow \mu_\uparrow + L_{22}^\downarrow \mu_\downarrow)],\nonumber \\
    C_s &=&  [(L_{21}^\uparrow - L_{21}^\downarrow) - (L_{22}^\uparrow \mu_\uparrow - L_{22}^\downarrow \mu_\downarrow)].
\end{eqnarray}

Obviously, the thermodynamical forces resulting from the gradients of temperature, magnetization, chemical potential etc. drive various coupled currents. In particular coupled currents result from $\Delta M(t,\phi,..)$. Note, $\Delta \mu_\sigma \sim \Delta H_{eff} +...
\sim \Delta M + ...$.  Clearly, the electron (charge) currents carry energy characterized by $\Pi_\sigma$ and also in general charge (if electron charge concentration gradients occur) and in magnetic systems spin polarization. The equations suggest that the spin voltage $\Delta (\mu_\uparrow - \mu_\downarrow)$ is affected by $\Delta T$ etc., see for example various stationary states (Seebeck effect).

Above equations allow to exploit the symmetries of the Onsager coefficients $L_{ij}$. For example, consider dependence on external
(magnetic) fields, since these may change the spin polarisation.

It might be useful to express the Onsager coefficients by the spin dependent transport coefficients, the electrical conductivity $\sigma_\sigma$, by the thermal conductivity
$\kappa_\sigma$, etc. and then to put the Onsager equations into the form \cite{1,2,4,5,9}
\begin{eqnarray}
        j_E &=& \Pi_\uparrow j_\uparrow + \Pi_\downarrow j_\downarrow
            = - (1/e)[\sigma_\uparrow \Pi_\uparrow \Delta \mu_\uparrow +
              \sigma_\downarrow \Pi_\downarrow \Delta \mu_\downarrow] + \kappa \Delta T + ...\quad ,\nonumber \\
\nonumber \\    j_e &=& - (1/e)[ \sigma_\uparrow \Delta \mu_\uparrow + \sigma_\downarrow \Delta \mu_\downarrow] + [\sigma_\uparrow S_\uparrow +  \sigma_\downarrow S_\downarrow ] \Delta T + ... \quad ,\nonumber \\
        j_\uparrow &=& - \sigma_\uparrow [ (1/e) \Delta \mu_\uparrow + S_\uparrow \Delta T + ...]  \quad,
        j_s = j_\uparrow - j_\downarrow  .
\end{eqnarray}
Note, $j_s = j_\uparrow - j_\downarrow$. Here, $\Pi_\sigma = \frac{j_E}{j_\sigma}$ are the Peltier coefficients, already introduced before.

The spin dependent Seebeck coefficients $S_\sigma = (1/e) \Delta \mu_\sigma / \Delta T$ and Peltier coefficients $\Pi_\sigma$ are important parameters describing the spin dependent thermoelectric and thermomagnetic effects. The Eqs. above have been used also in previous studies referring not explicitly to Onsager theory, see MacDonald, Maekawa, Uchida, Slachter {\it et al.} \cite{1,4,5}.

In ferromagnets one has in general for the electrical conductivity $\sigma_\uparrow \neq \sigma_\downarrow$ and
$j_s = j_\uparrow - j_\downarrow \neq 0$ results already from the electric potential difference $\Delta \varphi$ alone, since the electron density of states (DOS) is spin dependent,
($N_\sigma(\varepsilon)$).

The important spin dependent Seebeck coefficients \cite{9}
\begin{equation}
  S_{\uparrow(\downarrow)} = 1/e (\frac{\Delta \mu_{\uparrow(\downarrow)}}{\Delta T}),
\end{equation}
express that $\Delta T$ induces a spin voltage contribution, see previous discussion \cite{1,4}. Note then, this gives
\begin{equation}
               \Delta (\mu_\uparrow - \mu_\downarrow) = e S_s \Delta T +...,
               \quad S_s = 1/e \frac{\Delta (\mu_\uparrow - \mu_\downarrow)}{\Delta T}.
\end{equation}
This is large in magnetic metals, if the DOS difference ($N_\uparrow(\varepsilon) - N_\downarrow(\varepsilon)$) is large for energies $\varepsilon$
around the Fermi--energy.

The spin Seebeck effect could be observed for the tunnel junction shown in Fig. 1. A current flows through two ferromagnets with temperature T and ($T+\Delta T$). Then a spin voltage is generated at the interface between FM 1 and 2 (acting like a condensator) for $j_s=0$.

Of course, in magnetic tunnel junctions the usual Seebeck coefficient \cite{9}
\begin{equation}
  S = \frac{\Delta \varphi}{\Delta T}
\end{equation}
depends on
the magnetic configuration of the tunnel junction, see Figs 1,2 and note giant magnetoresistance (GMR) or tunnel magnetoresistance (TMR). Then,
\begin{equation}
             \Delta S = S_{\uparrow \uparrow} - S_{\uparrow \downarrow}
\end{equation}
reflects this and gives the change of the Seebeck effect upon changing the magnetizations on the left and right side of the tunnel junction from $\uparrow \uparrow$ to antiprallel configuration $\uparrow \downarrow$.

More and detailed experimental studies are needed to determine Onsager coefficients, to check on previous equations, and to determine
different Onsager coefficients in external magnetic fields. The Onsager Eqs. show again that in particular the Josephson spin current due to $\Delta M$, with respect to its phase gradient, is accompanied by corresponding contributions to $j_e$, $j_E$, etc..

To repeat, the Onsager equations demonstrate
that the spin voltage gradient $\Delta(\mu_\uparrow - \mu_\downarrow)$, $\Delta T$, $\Delta M$, phase gradients drive coupled currents. Spin polarizations result from the molecular field $H_{eff}$ and local magnetic moments. Obviously, the currents are generally accompanied by energy flow
(heat flow). The charge current $j_e$ and spin current $j_s$, $j_\sigma$ may decouple.

Furthermore, for example current $j_E$ etc. may induce temperature gradient $\Delta T$ and thus a magnetization gradient $\Delta M$ may result. An external magnetic field may change the magnetization and thus affect the coupled currents.

Regarding spin currents it is important to note the general formula
\begin{equation}
             j_i  \propto  \Delta H_{eff} + ... \propto \Delta M(r, t) + ...,
\end{equation}
expressing that the magnetization gradient $\Delta M$, including in particular the gradient of the phase of the magnetization, drives various coupled currents ( for example in a tunnel junction ).

It is of interest to analyze, the Onsager equations, the currents for special situations.

\subsection{Stationary State $j_e = 0$, $j_s = 0$}

It follows from the above equation for vanishing charge current
$j_e = 0$ :

(a)

\begin{equation}
   C_e \Delta T \simeq - T [L_{22}^\uparrow \Delta (\mu_\uparrow + \mu_\downarrow) + (L_{22}^\downarrow - L_{22}^\uparrow) \Delta
   \mu_\downarrow ].
\end{equation}
Then, neglecting last term and assuming $L_{22}^\uparrow \approx L_{22}^\downarrow$ one gets ($\Delta ( \mu_\uparrow + \mu_\downarrow)
\propto e \Delta \varphi$)
\begin{equation}
         \frac{\Delta \varphi}{\Delta T} = \frac{C_e }{2Te L_{22}^\uparrow}.
\end{equation}
Here, $S = (1/e) \Delta \varphi / \Delta T$ is the previously defined Seebeck coefficient \cite{9}.

(b) $j_s = 0$ :

For vanishing spin current one finds similarly
\begin{equation}
        S_s \equiv  \frac{\Delta (\mu_\uparrow - \mu_\downarrow)}{\Delta T} \simeq - \frac{C_s}{T L_{22}^\uparrow},
\end{equation}
where $S_s$ is the spin Seebeck coefficient. Note, $S_\sigma = (1/e) \frac{\Delta \mu_\sigma}{\Delta T}$ and $S_s = S_\uparrow - S_\downarrow$, and
$S = S_\uparrow + S_\downarrow$.

Obviously, the temperature gradient $\Delta T$ generates the gradient
$\Delta ( \mu_\uparrow - \mu_\downarrow )$ of the spin voltage (or vice versa).
The Seebeck effect applies, for example, to magnetic film layers or tunnel junctions ($F_1/N/F_2$) with temperature gradient or gradient of the magnetization.

The usual Seebeck coefficient S, see de Groot \cite{9}, reflects of course like magnetoresistance the magnetic configuration in film layers or tunnel junctions ($F_1/.../F_2$), ($S_{\uparrow\uparrow} \neq S_{\uparrow\downarrow}$ in general, where $\uparrow \uparrow$ refers to parallel and $\uparrow \downarrow$ refers to antiparallel magnetization of $F_1$ and $F_2$, respectively, see Figs. 1 and 2.

An external magnetic field $\overrightarrow{H}$ affects the spin voltage. Regarding the spatial dependence $x$ of the spin dependent chemical potential $\mu_\sigma(x,t)$, note for $\frac{dT}{dx}= const.$ the gradient of the spin voltage varies linearly for a (one--dim.) tunnel junction in $x$--direction.

The spin Seebeck effect means as discussed before that a spin voltage can be induced in a magnetic metal without an electric current ($\Delta \varphi = 0$), since
$\Delta T$ causes a contribution to the spin voltage $\Delta (\mu_\uparrow - \mu_\downarrow) \neq 0$.

The spin current $j_s$ is expressed by $L_{ij}^\sigma$ and approximately $j_s \propto ( N_\uparrow (\varepsilon_F) - N_\downarrow (\varepsilon_F) )$ + ...
As is clear this spin current depends on the spin mean free path, but might disappear due to spin--flip scattering ( see for comparison spin currents injected into metals ). For example in a tunnel junction involving tunneling through graphene (with spin dissipation length $\sim$ nm or more) one might get relatively large spin currents induced by a temperature gradient. This is also the case for the spin currents resulting from the gradient of the phase of the magnetization.

Regarding dynamics of currents the time dependence of the gradient of
the magnetization phase is of interest.

Note, in a ferromagnet at nonequilibrium with hot electrons the chemical potential might change in time t and then $\mu_\sigma(x,t,...)$. This is expected to yield interesting dynamical behavior.

\subsection{Heat Transport due to Spin Currents ($\Delta T = 0$, $\Delta \varphi \neq 0$).}

As is evident from the analysis above the electronic currents carry energy in particular also the spin currents. The Peltier coefficients $\Pi_\sigma$
describe this. Note,
\begin{equation}
   \left.  \frac{j_E}{j_e}\right)_{\Delta T=0} = \Pi  ,  \quad
   j_E = \Pi_\uparrow j_\uparrow + \Pi_\downarrow j_\downarrow \quad,
   \Pi_\sigma = \frac{j_{E\sigma}}{j_\sigma}.
\end{equation}

Special cases:

(a) $\Delta T = 0$, $\Delta \varphi \neq 0$.

Then it is approximately
\begin{eqnarray}
       j_E\!\!\!\! &=& \!\!\!\!  (1/T) \Sigma_\sigma L_{12}^\sigma \Delta \mu_\sigma +...
           \approx (1/T) L_{12}^\uparrow \Delta (\mu_\uparrow + \mu_\downarrow) +... ,\nonumber \\
       j_e\!\!\!\! &=&\!\!\!\! (1/T) \Sigma_\sigma L_{22}^\sigma \Delta \mu_\sigma +...
       \approx  (1/T) [ L_{22}^\uparrow \Delta (\mu_\uparrow + \mu_\downarrow) + ( L_{22}^\downarrow - L_{22}^\uparrow ) \Delta \mu_\downarrow ]       ...,\nonumber \\
       j_\uparrow\!\!\!\! &=& \!\!\!\!(L_{21}^\uparrow- L_{22}^\uparrow \mu_\downarrow)\frac{\Delta T}{T^2}+ (1/T) L_{22}^\uparrow \Delta \mu_\uparrow + ... ,\nonumber \\
       j_s\!\!\!\! &\simeq&\!\!\!\! (1/T) L_{22}^\uparrow \Delta (\mu_\uparrow\!\!-\! \mu_\downarrow) +... \; .
\end{eqnarray}

Hence, for $\Delta T=0$ it is $j_E = \Pi j_e$ and
\begin{equation}
       \Pi = \frac{j_E }{j_e} \approx \frac{L_{12}^\uparrow}{L_{22}^\uparrow}+... .
\end{equation}
Then the Seebeck coefficient is given by $S = - \frac{\Pi}{T}$  \cite{9}. Also approximately
\begin{equation}
       \Delta \mu_\uparrow \simeq - \frac{\Pi_\uparrow}{T}+...,\nonumber \\
       \frac{\Delta (\mu_\uparrow - \mu_\downarrow)}{\Delta T} \simeq -(1/T) [( \Pi_\uparrow - \Pi_\downarrow) -
                  (L_{22}^\downarrow - L_{22}^\uparrow)/ L_{22}^\uparrow ) \Pi_\downarrow  +...] .
\end{equation}
Note, one gets approximately $j_{E\uparrow} = (1/T) L_{12}^\uparrow \Delta \mu_\uparrow +...$, and $j_\uparrow = (1/T) L_{22}^\uparrow \Delta \mu_\uparrow +...$. Thus
\begin{equation}
      \Pi_\uparrow = \frac{j_{E\uparrow}}{j_\uparrow} \approx \frac{L_{12}^\uparrow}{L_{22}^\uparrow}+ ...
\end{equation}
Then,
\begin{equation}
        \frac{\Delta (\mu_\uparrow - \mu_\downarrow)}{\Delta T}\approx - \frac{\Pi_\uparrow - \Pi_\downarrow}{T}+... .
\end{equation}

Charge and spin currents generate heat in a magnetic tunnel junction which affects gradients in the magnetization.

For a tunnel junction with a magnetic metal A on the left side and a metal B on the right side, see Fig. 3, one gets in the tunnel medium heat generation
\begin{equation}
             j_E^A - j_E^B = ( \Pi_{i}^A - \Pi_{i}^B ) j_i  ,
\end{equation}
where i refers to an electron current ($i=e$), current for electrons with spin $\sigma$ ($i=\sigma$) and spin current ($i=s$). The generated heat $\Pi_{AB}$  ($\Pi_{AB} = \Pi^A - \Pi^B$) is observable in particular if the tunnel medium between A and B is magnetic, for example ferromagnetic, or is superconducting. Note, besides heat also radiation may occur at the contact of A and B. A special situation is if $j_e^A \approx j_\uparrow$ and $j_e^B \approx j_\downarrow$.

Also in accordance with magnetoresistance (GMR or TMR) the current discontinuity $\Delta j_E = (j_E^A - j_E^B)$
for electron currents $i=e$ (due to the spin dependent electron conductivity $\sigma_\sigma$) depends on the direction of the magnetizations of metals A and B.

Of interest is to study the Seebeck effect for a tunnel junction consisting of three ferromagnets in series, see Fig.2, and
to observe the dependence of the spin voltage $\mu_\uparrow - \mu_\downarrow$ on the magnetic configuration of the three ferromagnets, e.g. a.f. vs. ferromagnetic one. Clearly, operating two tunnel junctions as parallel circuits may yield interesting interference effects of the currents.

One may express the heat and spin current within an electronic theory and obtains thus an expression for the spin dependence of the Peltier heat suitable for an electronic calculation. It needs to be studied how characteristically the Peltier heat depends on the ferromagnet, its magnetisation.

In Fig. 3 the Peltier effect is sketched.
\begin{figure}
\centerline{\includegraphics[width=.6\textwidth]{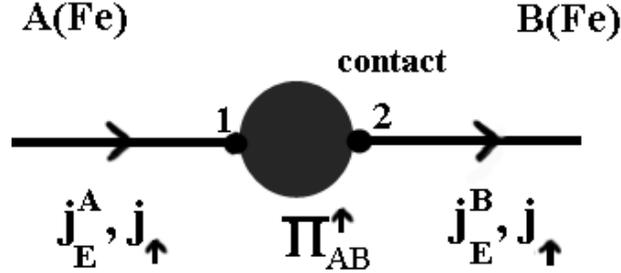}}
\caption {Peltier effect for two ferromagnets A, B (for example Fe). Spin dependent heat $\Pi_{\sigma}$ develops at contact, for example Cu or a magnetic transition metal, due to spin current $j_\sigma$. Contact 1 may warm up and 2 cool down, and heat $P=\Pi j$ occurs which should depend on spin polarisation, on $(j_\uparrow-j_\downarrow)$. Of course, the Peltier heat will also depend on the magnetic configuration of the magnets A and B and is different for ferromagnetic and antiferromagnetic arrangement of the two magnets. Note, in case of two contacts in sequence and ferromagnets A, B, C, and in series interesting interferences may occur.}
\end{figure}
As indicated by the Fig.3 in particular the Peltier heat of itinerant ferromagnets is expected to depend on the relative magnetisations of the ferromagnet and of course on the metals A and B.

As is clear from Fig. 1 and from Onsager Eqs. (for $\Delta \varphi = 0$) the temperature gradient $\Delta T$ induced by the currents will affect
the magnetization $\Delta M = M(T_1) - M(T_2)$ and thus change the current driven by $\Delta M$. (Approximately one gets from Onsager equations
$\Delta T \simeq \Delta (\mu_\uparrow - \mu_\downarrow) / S_s$).

Regarding magnetic nanostructures, note the system sketched in Fig.2 may yield as mentioned already oscillating currents $j(t)$ and $j_\sigma(t)$ and which are optically induced.

Creating optically for example in the ferromagnetic metal 2 hot electrons then the magnetization $M_3$ decreases by $\Delta M(t)$
\cite{13}.
This changes the magnetoresistance and affects the currents $j_e$ and $j_\sigma$. After ultrafast relaxation of the hot electrons one may repeat the excitation of the electrons. This yields the ultrafast oscillations of the currents. Thus one may also manipulate the Kirchhoff emission \cite{2}.

Of course, one may also apply the Onsager theory to currents in ring structures with gradient forces $X_{i\sigma}$ to obtain interesting thermoelectric and thermomagnetic effects, including Aharonov--Bohm effect etc.\cite{8,2}.

\subsection{Tunnel Junctions involving Superconductors}

Of interest is also to use superconductors as a spin filter, see illustration in Fig.4 \cite{10}. As known a singulet superconductor may
\begin{figure}
\centerline{\includegraphics[width=.5\textwidth]{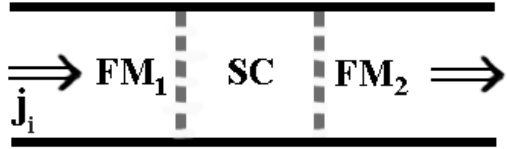}}
\caption{Illustration of a magnetic tunnel junction consisting of two ferromagnetic metals $(FM)_1$ and $(FM)_2$ separated by a superconductor (SC).
The electron current $j_e$ as well as $j_\uparrow$ and $j_\downarrow$ and $j_s=j_\uparrow - j_\downarrow$ depend on
the relative orientation of the magnetisations $\protect\overrightarrow{M_1}$ and $\protect\overrightarrow{M_2}$ and on the superconducting state,
singulet vs. triplet Cooper pairs. Note, $M_2 - M_1 = \Delta M$ may cause Josephson like spin current which is particularly affected in case of a triplet superconductor by the phase of its order parameter. Of course, the spin current is destructively affected by spin flip scattering.}
\end{figure}
block a spin current and affect the currents driven for example by the gradients $\Delta T$, $\Delta M = M_1 - M_2$, etc.. Depending on the energy gain due to $j_e$ vs. loss of energy due to (singulet) Cooper pair breaking one may get that the currents weaken the superconducting state. Note, $\Delta M(t)$ may cause Josephson like spin current ($j_s \propto \sin \Delta \phi + ...$) \cite{12}.

If the two ferromagnets are separated by a triplet superconductor, then the relative orientation of the angular momentum $\overrightarrow{d}$ of the
triplet Cooper pairs with respect to the magnetizations $\overrightarrow{M_1}$ and $\overrightarrow{M_2}$ controls the tunnel currents \cite{2,7,12}.

Note, $\overrightarrow{d}$ may be oriented via an external magnetic field.
\begin{figure}
\centerline{\includegraphics[width=.5\textwidth]{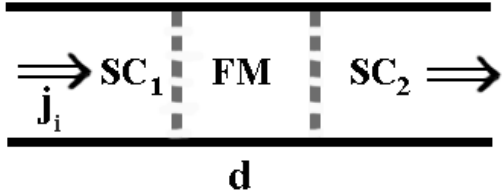}}
\caption{Tunnel current $j$ between two superconductors $SC_1$ and $SC_2$ which depends on the relative phase of the order parameter of the two superconductors and the phase of the magnetization. Of course, the thickness d of the ferromagnet controls the current, in particular spin polarized ones. The tunnel current may be manipulated optically via hot eletrons in the ferromagnet.}
\end{figure}
Of particular interest is to study the effect of superconductivity, triplet superconductivity on the (giant) magnetoresistance in case of two antiferromagnetically (af) oriented ferromagnets, see Fig.4. One expects for parallel configuration of $\overrightarrow{d}$, $\overrightarrow{M_1}$ and $\overrightarrow{M_2}$ the lowest resistance, while the largest one for af configuration of $\overrightarrow{d}$, and magnetizations. Of course
Onsager theory can be used to describe the system illustrated in Fig.4. Note, $M_2-M_1$ may act like a magnetization gradient.

It is of general interest to test tunnel junctions like shown in Fig. 5 with respect to acting as filters for longitudinal vs transverse currents.
Spin orbit coupling, fields H and $H_{eff}$ etc. will control this.

Related properties are expected for the tunnel junction shown in Fig.5. One may use this to distinguish singlet from triplet superconductivity. Onsager theory can be used to describe such a system phenomenologically. Josephson currents $j_J$ characterize sensitively such tunnel junctions. The current $j_J$ decreases for increasing thickness d of the ferromagnet and for decreasing Cooper pair binding energy ($T_c$). Also in the spirit of Onsager theory the difference ($\Delta_2 - \Delta_1$) of the
superconducting order parameters acts like a gradient inducing corresponding currents.

In case of triplet superconductivity (TSC) the Josephson current $j_J$ depends in an interesting way on $T_c$ and the angle between the magnetization $\overrightarrow{M}$ and direction normal to $\overrightarrow{j_J}$. The current should depend on the triplet state and
impurity scattering (in particular spin orbit scattering). Hot electrons in the ferromagnet FM modulate $j_J$. Generally the spin polarization of
the currents may be manipulated by the gradient $\Delta M(t)$.

In view of the significance of occurrence of triplet superconductivity in metals we sketch the situation in the following Fig 6. The current carried by Andreev states is
calculated using \cite{7,12}
\begin{equation}
         j_J = - (e/\hbar) \sum \frac{\partial E_i}{\partial \phi} \tanh (E_i / 2kT),
\end{equation}
where over all Andreev states with energy $E_i$ and mediating the tunneling is summed. Here, $\phi$ is the phase difference between the
Cooper condensates on the left and right side of the tunnel junction. One expects $j_J$ to depend characteristically on the phases
of all order parameters, on the relative orientation of the Cooper pair vectors $\overrightarrow{d_L}$, $\overrightarrow{d_R}$ and magnetization  $\overrightarrow{M}$, respectively. The triplet Cooper pairs are described by $\Delta (k) = \sum_l d_{l}(k) (\sigma_l i \sigma_2)$, l = 1,2,3
where $\sigma_l$ are the Pauli spin matrices and $d_l$ are the spin components of the superconducting order parameter, see Bennemann and  Ketterson \cite{7,12}. Note, the triplet Cooper pairs have a spin and orbital momentum.

For the transport of angular momentum, obviously the phases of all three order parameters are of importance for tunneling. Even for no phase difference $\phi = \phi_L - \phi_R$ between the
triplet Cooper pair condensates on both sides of the tunnel junction
one gets for arbitrary phase of the magnetization of the ferromagnet a Josephson current.
In the ferromagnet the Andreev states carry the current of the tunneling electrons and temperature controls its population.
Also, of course, the magnitude of the magnetization and electron spin relaxation in the FM matter.
\begin{figure}
\centerline{\includegraphics[width=.6\textwidth]{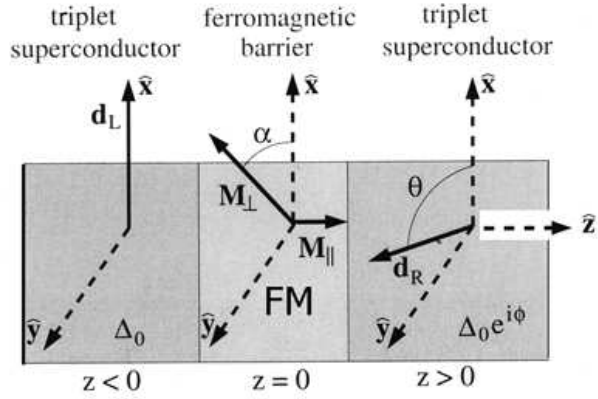}}
\caption{Illustration of a tunnel junction ($TSC/FM/TSC$). The phases $\Theta$, $\phi$ and $\alpha$ of the superconducting order parameter, Cooper pair condensate and of the magnetization $\protect\overrightarrow{M}$, respectively, control the tunnel currents. The magnetization may be decomposed into components $M_{\bot}$ and $M_{\parallel}$. Due to the spin and angular momentum of the Cooper pairs one expects that the current through the FM depends sensitively on the relative direction of $\protect\overrightarrow{M}$, spin relaxation, spin flip scattering resulting for example from spin--orbit coupling, population of the Andreev states and thickness d of FM.}
\end{figure}

As physically expected the Josephson current may change sensitively upon rotation of $M_\bot$, change of the direction of $\overrightarrow{M}$. Model calculations yield
results shown in Fig.7(a) \cite{7,12}. This implies that the tunnel junction (TSC|FM|TSC) may act like a switch turning on and off the current $j_J$. This behavior suggests a sensitive dependence of the current $j_J$ on an external magnetic field.

In Fig.7(b) model calculation results, simplifying strongly the influence of the FM metal, are given for the temperature dependence of the Josephson current \cite{7,12}. These should reflect the temperature controlled occupation of the Andreev states. The change in sign of $j_J(T)$ as a function of T
occurs only if Andreev states are non degenerate and in case of two Andreev states 1 and 2 which derivatives
$\frac{\partial E_1}{\partial \phi}$ and $\frac{\partial E_2}{\partial \phi}$ have opposite sign. Also note, the sign change of $j_J$
for increasing temperature may be suppressed by electron scattering in the FM \cite{7,12}.

Clearly, in view of the importance studying triplet superconductivity improved calculations of the current $j_J$ are needed. The FM tunnel junction
metal must be taken into account in a more realistic way.

\begin{figure}
\centerline{\includegraphics[width=.7\textwidth]{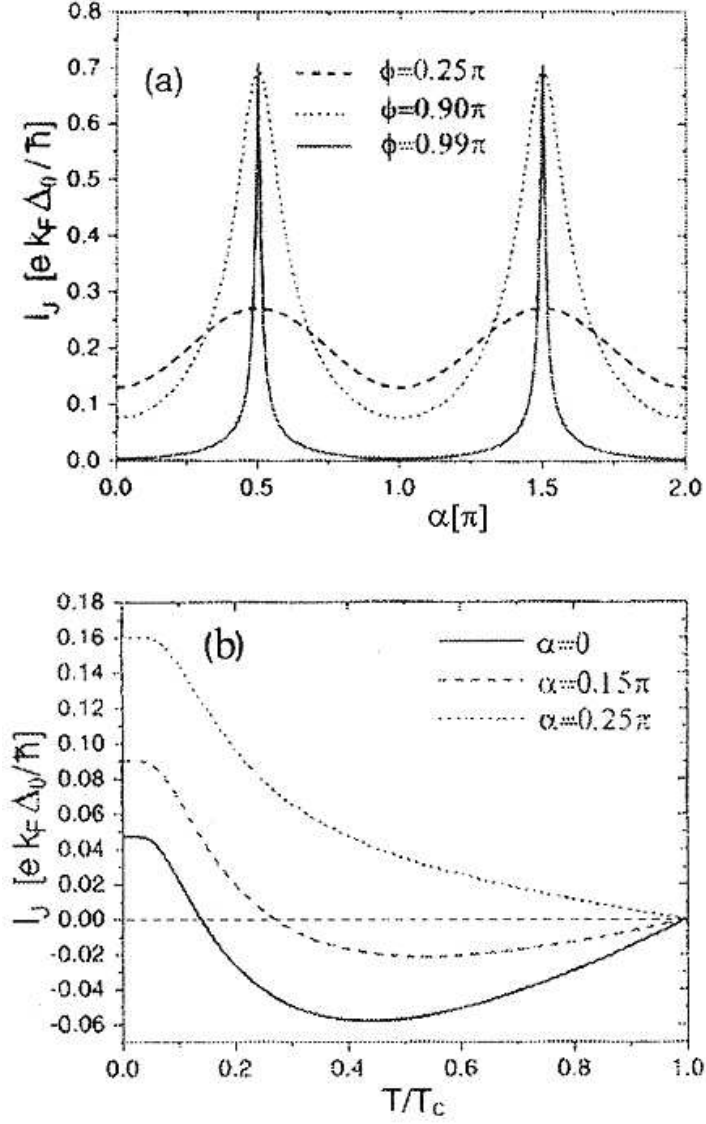}}
\caption{Results for a tunnel junction (TSC/FM/TSC) sketched in Fig.6, see Morr {\it et al.}. Dependence of the Josephson current $I_J$ on (a) phase $\alpha$ of the magnetization $\protect\overrightarrow{M}$ and (b) temperature T for various values of $\alpha$. Results refer to model calculations simplifying the coupling of the Cooper pairs to the FM. For increasing electron scattering at the ferromagnetic barrier
$j_J$ does not change sign for increasing temperature any more. The Andreev states carrying the current are determined using Bogoliubov--de Gennes method. $T_c$ is the superconducting transition temperature.}
\end{figure}

In case of paramagnetism and $M=\chi H$, where $\chi$ denotes the spin susceptibility, the direction of the external magnetic field can be used to manipulate the current.

In view of this rich behavior of (TSC/FM/TSC) tunnel junctions one expects also interesting behavior for the currents through junctions (FM/TSC/FM). Again the phases of the three order parameters control the currents. One expects
\begin{equation}
            j_s^J \simeq A(\Theta) \sin(\Delta \phi + \eta),
\end{equation}
where $\Theta$ refers to the relative phase of the triplet Cooper pairs and $\Delta \phi$ to the phase difference of the magnetization on L--side and right--side of the junction. Weak spin--orbit scattering and long spin free--path for tunneling in the TSC favor the current.

Of course, the Josephson spin current $j_s^J$ can also be manipulated optically by changing the population of the electronic states, by an external
magnetic field and by applying a temperature (or pressure) gradient to the tunnel junction.

A spin current $j_s=j_\uparrow - j_\downarrow$ may result due to $N_{\sigma L}(\varepsilon) \neq N_{\sigma R}(\varepsilon)$ for the electronic density of states. Then, $j= j_s + j_s^J+...$.

\subsection{Galvanomagnetic Effects}

Extending as usually the Onsager theory in the presence of an external magnetic field $\overrightarrow{H}$ and $\overrightarrow{M}$ causing space anisotropy one obtains the spin dependent galvanomagnetic effects, the Hall--effect, Nernst--effect, etc.. Then one gets currents
\begin{equation}
            j_{i}^x  , j_{i}^y
\end{equation}
and in particular currents $j_e^x, j_e^y, j_E^x, j_E^y, j_{\sigma}^x, j_{\sigma}^y$, etc. due to forces $X_{e}^\alpha$, $X_{E}^\alpha$, etc..

For illustration see Fig.8.
\begin{figure}
\centerline{\includegraphics[width=.5\textwidth]{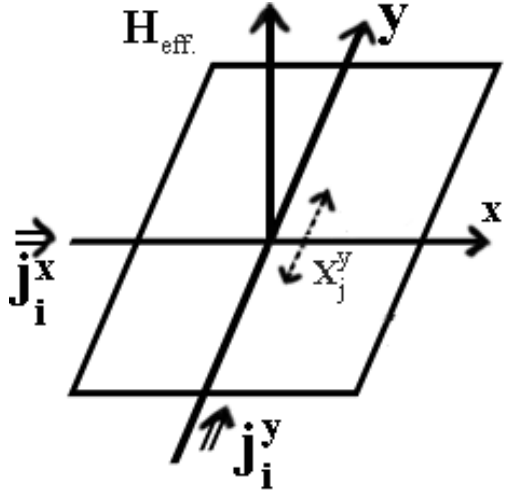}}
\caption{Currents $j_i^{\alpha}$ are driven by gradients $\Delta T^\alpha$, $\Delta \varphi^\alpha$, $\Delta \mu_\sigma^\alpha$, $\Delta M^\alpha$, etc. and depend on the relative orientation of the external magnetic field $H$ ($H \bot x,y$ plane) and magnetization $\protect\overrightarrow{M}$ which induce spatial anisotropy ($\alpha = x,y$). Indicated is the spin dependent gradient $X_{j}^y$ induced by currents $j_e^x$, $j_\sigma^x$, etc.. The index $j$ may refer to $j=E$, $\varphi$, $\sigma$, $s$, $\mu_\sigma$, etc. Note, if the magnetization points perpendicular to the $x,y$--plane, then $M$ may act like $H$ and one gets also for $H=0$ a Hall potential. For simplicity the magnetization $\protect\overrightarrow{M}$ is taken to be in the ($x,y$)--plane parallel to $x$ or perpendicular to the ($x,y$)--plane. For example the current$j_{s}^x$ induces a spin voltage $\Delta_{y}(\mu_\uparrow - \mu_\downarrow)$ and $j_{e}^x$  a voltage $\Delta_{y} \varphi$, $j_{E}^x$ and gradient $\Delta_{x} T$ may also cause a spin voltage in y--direction (Nernst--effect). Spin--orbit scattering affects $\Delta_{x(y)} \mu_\sigma$. In accordance with the Lorentz force  (in presence of a molecular field $H^{'}$) one expects also for 2d--structures (graphene, etc.) large Hall--effects. Currents should depend sensitively on hot electrons (for example due to light). To simplify the analysis one may assume symmetry with respect to $x\protect\rightleftarrows y$ and also for Onsager coefficients $L_{ij}(H_{eff})=L_{ji}(-H_{eff})$, etc..}
\end{figure}
Note, if M $\bot x,y$--plane, then one gets also for no external magnetic field, $H=0$, a Hall--effect due to temperature gradient $\Delta T$, electric potential gradient $\Delta \varphi$ , magnetization gradient $\Delta M(t)$, etc..

To determine the spin dependent Thermomagnetic and Galvanomagnetic effects due to the anisotropy resulting from external magnetic field or $\overrightarrow{H_{eff}}$ one may use ( for $X_E^\alpha = - (1/T^2) \Delta_\alpha T$,
$X_{i\sigma}^\alpha = \Delta_\alpha (\frac{\widetilde{\mu_\sigma}}{T}) = \Delta_\alpha [(1/T) ( -e \varphi + ( \mu(0)-\sigma \mu_0 H_{eff}) ]$,
etc and $\alpha = x, y$ ) Onsager equations for
$j_{E}^x$, $j_{E}^y$, $j_{e}^x$, $j_{e}^y$, etc.. Thus, one may write
\begin{eqnarray}
 j_{e\sigma}^x &=& L_{11}^\sigma X_\sigma^x + L_{12}^\sigma X_\sigma^y + L_{13}X_E^x + L_{14} X_E^y + ... ,\nonumber \\
 j_{e\sigma}^y &=& L_{21}^\sigma X_\sigma^y + L_{22}^\sigma X_\sigma^y + L_{13}X_E^x + L_{14} X_E^y + ... ,\nonumber \\
 j_E^x &=& L_{31}^\sigma X_\sigma^x  + L_{32}^\sigma X _\sigma^y + L_{33} X_E^x + L_{34} X_E^y + ... ,\nonumber \\
 j_E^y &=& L_{41}^\sigma X_\sigma^x  + L_{42}^\sigma X _\sigma^y + L_{43} X_E^x + L_{44} X_E^y + ... ,\nonumber \\
 j_s^x &=& j_{e\uparrow}^x - j_{e\downarrow}^x  ,\nonumber \\
 j_s^y &=& j_{e\uparrow}^y - j_{e\downarrow}^y  .
\end{eqnarray}
($j_e^x = j_{e\uparrow}^x + j_{e\downarrow}^x$, etc.). It is straightforward to write explicitly all terms in the Onsager equations. It is important to note that the Onsager equations yield coupling of all quantities and in particular coupled currents $j_{i}^x$ and
$j_{i}^y$ driven by the forces $X_{l}^x$ and $X_{l}^y$, see de Groot \cite{9}. Symmetry like isotropy with respect to
$x \leftrightarrows y$, $L_{ij}(H) = L_{ji}(-H)$, $j_e^x\rightarrow - j_e^x$ for $H\rightarrow -H$, $X_i \rightarrow -X_i$ etc. reduce the number of Onsager coefficients to (about) 12 (due to spin dependence 6x2) which can be expressed by transport coefficients  ($\sigma_\sigma^\alpha$, etc.).
Note, for convenience we write $L_{ij}^\alpha \rightarrow L_{ij}$ and use different indices i,j for $\alpha = x$ and $\alpha = y$.

Due to linearity the Onsager Eqs. can be rewritten by putting the experimentally controlled quantities on the right side of the equations, $X_\sigma^\alpha \rightleftarrows j_\sigma^\alpha$, see de Groot \cite{9}. Thus,
\begin{equation}
  \Delta_x X_\sigma^x = \widetilde{L_{11}^\sigma} j_{e\sigma}^x + \widetilde{L_{12}^\sigma
  } j_{e\sigma}^y + L_{13} X_E^x + L_{14} X_E + ...
\end{equation}
and similar equation for $ \Delta_y X_\sigma^y = ...$, and as before $j_E^\alpha = ...$. Thus, one gets $X_\sigma^\alpha = \Delta_\alpha \frac{\widetilde{\mu_\sigma}}{T}$ :

\noindent {\bf(1)} The Hall--effect, when the current $j_{s}$ in x--direction generates a spin voltage in y--direction. It is ( spin Hall--effect )
\begin{equation}
                 \Delta_{y} (\mu_\uparrow - \mu_\downarrow) = a(H_{eff}) j_{s}^x + ... \quad    .
\end{equation}
Thus, the spin current in x--direction induces a spin voltage in y--direction. Also $\Delta_y X_\sigma^y \sim j_{e\sigma}^x +..$.
The analysis uses symmetry with respect to $x\rightleftarrows y$.
For $\overrightarrow{H_{eff}} \bot x,y-plane$ one gets using standard analysis
\begin{equation}
    a(H_{eff}) \propto H_{eff} + ... \quad .
\end{equation}

Note, if two magnetic metals are put together parallel to the x--direction in the x,y--plane, then one expects due to TMR (or GMR)--effects
a particularly large generation of a spin voltage in y--direction.

Of course, $j_{e}^x$ generates also $\Delta_y \varphi$ which depends on $H_{eff}$ and
\begin{equation}
        \Delta_y \varphi = \tilde{a}(H_{eff}) j_e^x + ...
\end{equation}
(usual Hall--effect).

\noindent {\bf(2)} Nernst--effect:
generation of a spin voltage in y--direction due to $j_{E}^x$ and $\Delta_{x} T$. It is ( spin Nernst--effect )
\begin{equation}
                 \Delta_{y} (\mu_\uparrow - \mu_\downarrow) = b(H_{eff}) \Delta_{x} T + ... \quad.
\end{equation}
Applying usual symmetry arguments if $H_{eff}$ is perpendicular to the x,y--plane one gets $b$ $\propto H_{eff}$. Interesting behavior may occur if
the external magnetic field H and
the molecular field $qM$ are not collinear.

\noindent {\bf(3)} The other effects like Ettinghausen one ( $j_e^x \rightarrow \Delta_y T$, for $j_E^y = j_E^y =0$,  etc. ) are obtained similarly from the Onsager equations, see Fig.8 for illustration.

In the equations above we neglected for simplicity terms
due to the interdependence of the currents $j_{e}^\alpha$ and $j_{s}^\alpha$.

Summary : The various effects (currents) arising for forces $T X_i^\alpha = - \Delta_\alpha \widetilde{\mu_i}$, where $\widetilde{\mu_i}= - e \varphi + \mu_\sigma$, $\mu_\sigma = \mu(0) - \sigma \mu_0 H_{eff}$, may be summarized by

\bigskip
$\left(
  \begin{array}{c}
     - \Delta_x \widetilde{\mu_\sigma} \\
      - \Delta_y \widetilde{\mu_\sigma} \\
     j_{E\sigma}^x  \\
        j_{E\sigma}^y  \\
  \end{array}
\right) =
\left(
  \begin{array}{cccc}
      \sigma_\sigma^{-1} &  H R_\sigma &   -\varepsilon_\sigma &   - H\eta_\sigma  \\
      -HR_\sigma  &  \sigma_\sigma^{-1} & H\eta_\sigma &  -\varepsilon_\sigma \\
     -T\varepsilon_\sigma & -TH\eta_\sigma   & -\kappa  &   - H\kappa L \\
    TH\eta_\sigma  & -T\varepsilon  &    H\kappa L &  -\kappa \\
  \end{array}
\right)
\left(
\begin{array}{c}
  j_\sigma^x \\
  j_\sigma^y \\
  \Delta_x T \\
 \Delta_y T
\end{array}
\right).$
\bigskip
Here, $L=\frac{\Delta_y T}{H \Delta_x T}$, thermoelectric power $\varepsilon_\sigma = -\frac{\Delta_x \widetilde{\mu_\sigma}}{\Delta_x T}$, Ettinghausen coefficient
$E_\sigma = \frac{\Delta_y T}{H j_\sigma^x}$, Hall coefficient $R_\sigma = - \frac{\Delta_y \widetilde{\mu_\sigma}}{H j_e^x}$ and
Nernst coefficient $\eta_\sigma = - \Delta_y \widetilde{\mu_\sigma}/ H \Delta_x T$.

As mentioned it is of interest to calculate the Onsager coefficients (transport coefficients) via the corresponding correlation functions \cite{14,15}.
Again, of interest are effects due to magnetization gradients. As remarked, new effects are expected due to a gradient in the phase of the
magnetization.

\subsection{Currents in Magnetic Rings}

Of special interest is to observe spin electron currents in magnetic rings (see persistent currents \cite{17}, and Aharonov--Bohm
effect), in optical lattices, or in magnetic quantum dot systems.

For magnetic rings, see Fig. 10 for illustration, interesting electronic structure may cause special behavior \cite{16,17,18}. The electron density of states (DOS), which for magnetic rings is spin dependent,
exhibits oscillations due to the interferences of the most important closed electron orbits (yielding the polygonal paths of the electron current)
of the ring, see Stampfli {\it et al.} \cite{16} and Fig.10. A magnetic field B inside the ring (and directed perpendicular to the ring) causing a flux,
\begin{figure}
\centerline{\includegraphics[width=.5\textwidth]{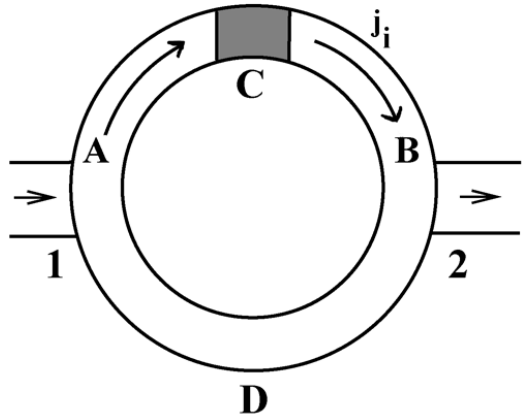}}
\caption{Illustration of Spin currents $j_i$ in a thin ring consisting of two magnetic metals A and B interrupted by a tunnel junction C. The spin dependent density of states $N_{\sigma}(\varepsilon)$ cause spin polarized currents. In case of contacting the ring at 1 and 2 with external force sources interesting interferences of the currents through A and D may occur. Currents may be driven by applying (spin dependent) gradients $X_{i}(t)$. Applying an external magnetic field perpendicular to the ring and inside the ring causes Bohm--Aharonov effect. Depending on the electron orbitals carrying the current interesting structure occurs for the density of states and
thus for the currents. Of particular interest are persistent (spin) currents arising for relatively large electron mean free path (comparable to the ring dimension) and which are stabilized, for example, by angular momentum conservation.}
\end{figure}
\begin{equation}
                  \phi = B S,
\end{equation}
where $S$ is the area enclosed by the electron orbits, yields via the Aharonov--Bohm effect ring currents (c = 1) \cite{19}
\begin{equation}
        j = - \frac{dF}{d\phi },
      j_s = j_\uparrow - j_\downarrow .
\end{equation}
A spin polarized current occurs in magnetic rings (also possibly in paramagnetic metals due to the spin polarization by an external field B).

Note, the free--energy can be written as $F = F_\uparrow + F_\downarrow$ and the spin dependent DOS yields
$F_\sigma$ \cite{19}.
\begin{figure}
\centerline{\includegraphics[width=.5\textwidth]{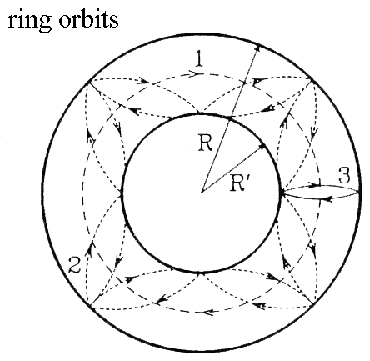}}
\caption{Most important electron orbits contributing within Balian--Bloch type theory to the electron structure of a ring and the electron DOS. Note, the electron orbits are deformed by the magnetic
field inside the ring. This modifies also the Aharonov--Bohm effect. The interferences of the electron paths within the ring cause oscillations in the DOS. The mean free--path of the electrons could be spin dependent causing interesting behavior.}
\end{figure}

A spin polarized current $j_s$ may also be induced by spin--orbit coupling \cite{18}. Note, $H_{so} \sim \overrightarrow{\sigma}\bullet \overrightarrow{L} \nabla V(r)$ where L is the angular momentum of the electrons in the field $V(r)$. The spin--orbit coupling causes
a phase $\phi_{AC}$ for electrons circling the ring and thus $j_{\sigma} \propto \frac{dF}{dt}$ determines the resultant current (Aharonov--Casher effect). This current may not be very small in case of strong spin--orbit coupling, for example for topological
insulators.

One gets from the theory by Stampfli {\it et al.}, which is an extension of the Balian--Bloch theory, that the electron density of states (DOS)
exhibits interesting structure, $N(\varepsilon, B)$, see Fig.11 for typical results. Then, ( for usual currents ) $j = j_\uparrow + j_\downarrow$, and approximately
$j_\sigma = \sigma_\sigma E$, and the conductivity is given by
$\sigma_\sigma = e^2 \tau_{\sigma} (\varepsilon)\overline{v v} N_{\sigma}(\varepsilon)$, with $\varepsilon \approx \varepsilon_F$.
Thus, one expects that the properties of the ring, (spin) currents etc. reflect the structure in the DOS.
\begin{figure}
\centerline{\includegraphics[width=.6\textwidth]{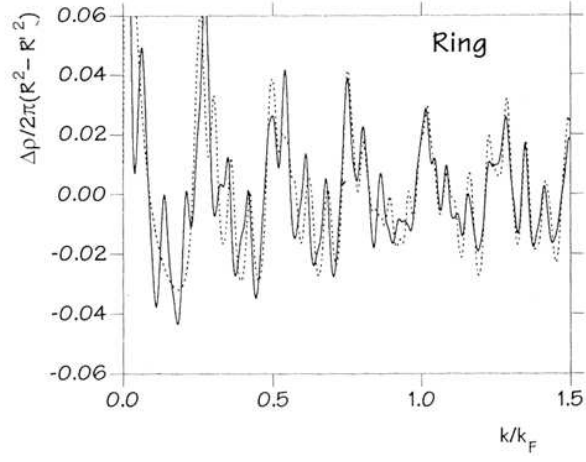}}
\caption{Electron Density of states (DOS) oscillations for a ring with outer radius R and inner radius $R^{'}$, $R' = 0.3 R$, using inside the
ring a square--well potential $U = - \infty$,
see Fig. 10. The dashed curve refers for comparison with Balian--Bloch type results to quantum mechanical calculations, see Stampfli {\it et al} \cite{16}. The DOS $\Delta \rho$ may be spin split by the molecular field $H_{eff}$.}
\end{figure}

The ring current driven by the magnetic flux (Aharonov--Bohm effect) is ($j\propto dF/dt$) approximately given by \cite{16,19}
\begin{equation}
     j = - \sum_{t,p,\sigma} \sin ( S B ) a_{t,p,\sigma}(B),
\end{equation}
where the coefficients $a_{t,p,\sigma}$ give the contribution to the current of the orbit characterized by t, p which are the numbers describing how many times the electron orbit circled the center of the ring and how many corners the polygon has, respectively. For illustration see the Figure 10.

The flux area depends due to the deformation of the polygonal paths on the magnetic field and $\phi = \pm B S_{\pm}, S_{\pm} = S_0 \pm \Delta$. Here, $\pm$
refers to clockwise (+) or counterclockwise (-) circling around center of the ring and $\Delta$ is the change of the area due to B \cite{16}.
Note, $\frac{dj}{dT}$ analysis or in general changing the parameters probing the electronic structure may exhibit the DOS structure.

One may also Fourier transform the free energy. The ring periodicity and invariance against time reversal and $B\rightarrow - B$, yields for $B=0$ inside ring as is the case for the superconducting state
$F = \sum_i F_i \cos (2\pi \phi)$ and thus
\begin{equation}
     j = \sum_{i \sigma} j_{i\sigma} \sin (2\pi i \phi/ (hc/e))
\end{equation}
for the current driven by the field B \cite{8}. Note, quantization of the flux in units of (hc/e) \cite{8}.

Regarding ring currents, in particular the persistent currents and Josephson type ones are interesting \cite{2,8,17}. These result if the mean free path of the current carrying electrons is large enough and comparable to the dimension of the ring and if spin relaxation is weak (weak spin--orbit coupling, ...). As discussed spin--orbit coupling yields also a flux $\phi_{AC}$ and thus persistent currents (Aharonov--Casher effect). This current is also expected to have structure due to the interference of the main electron orbits in the ring, see Stampfli {\it et al.} \cite{16}.

Tunnel junctions in rings display interesting behavior, for example of superconductivity, interplay of superconductivity and magnetism and of singulet and triplett Cooper pairing and $BSC \rightleftarrows BEC$ transition enforced by ring geometry.

Angular momentum conservation may stabilize such persistent currents, in particular spin polarized ones ($\sim \overrightarrow{j}\times \overrightarrow{R}$). From the results above, one may estimate $j \propto \frac{1}{R^n}$ for the currents of a ring with radius R and $n \geq 1$ (the ring current decrease is expected to depend essentially on $(l/R)$, where l is the average length of phase coherent electron motion). The temperature dependence is given
by the expression for the free--energy of the electrons \cite{19}.

Onsager theory may be used again to study currents driven by the gradients $X_{i}^\sigma$ acting (in addition to the Aharonov--Bohm effect) on the itinerant electrons in the ring. For example, the temperature at contacts 1 and 2, see Fig. 9, may be different and then the resultant gradient $\Delta T$ drives a current, similarly gradient
$\Delta \mu_{12,\sigma}$, etc.. One expects on general grounds that inhomogeneous magnetisation present in the ring induces as response spin currents. This is interesting if dissipation length gets comparable to ring length.

If the ring excludes the field B, for example by becoming superconducting, then as mentioned already the magnetic flux is quantized \cite{8}.

Tunneling through C from a magnetic metal A to a magnetic metal B depends as usual on the configuration of the magnetizations of A and B, see Fig. 9
for illustration. Thus via the giant magnetoresistance the current from 1 to 2  through metals A,C,B may be largely blocked in comparison to the one from 1 to 2 through magnetic
metal D ($j_1 = j_A + j_D$), spin dependently. Also note a spin current entering at 1 may flow (largely) through ACB if between 1 and 2 metal D
becomes superconducting. In general via a few parameters one may manipulate the interference of currents $j_A$ and $j_B$ and thus obtain interesting
behavior of the ring currents.

Again, a magnetic current $j_{AB} \propto \frac{dM}{dt} \propto \overrightarrow{M_A} \times \overrightarrow{H_{eff}}(B)$ occurs driven
by the magnetic phase gradient \cite{20}. This may yield interesting interferences.

Of interest is also to use a ring of a superconducting metal to study the transition $BSC \rightarrow BEC$ due to geometrical restrictions
(by changing the width of the ring, or narrowing (locally) the ring, etc, causing a corresponding change of the size, radius of the Cooper pairs versus distance between Cooper pairs, $n_s^{-1/3}$. Note, the coupling strength for Cooper pairing determines the size of the pairs. For Cooper pair size smaller than their distance on expects BEC--behavior \cite{21}).

This shows already the many possibilities to study interesting physics using metallic rings.

\subsection{Currents involving Quantum Dots}

\subsubsection{Quantum Dots}
The currents between magnetic quantum dots are illustrated in Figs.12 and13, see M.Garcia \cite{22}.
Spin dependent currents may result for quantum dots due to the Pauli principle and Coulomb interactions (see Hubbard hamiltonian) and in particular for magnetic quantum dots and magnetic reservoirs ($\mu_{i\sigma}$).

The spin dependent electron tunneling, hopping between the magnetic quantum dots, photon assisted, is described by the Hamiltonian
\begin{equation}
    H = \sum_{i\sigma} \varepsilon_{i\sigma} c_i^{+}c_i + \sum_{ij} T (c_i^{+} c_j + h.c.) + H_{R},
\end{equation}
where $\varepsilon_{i\sigma} = \varepsilon_{i\sigma}^0 + a(t) \cos(\omega t)$, T gives the electron hopping between quantum dots and $H_{R}$ describes the coupling to the metallic reservoirs. It is $\varepsilon_{i\sigma}^0(t) = \varepsilon_i^0 + Un_{i\overline{\sigma}}(t) + ...$. The time dependent coupling of the electrons to the photons is given by $a(t) \cos(\omega t)$. The equations of motion for the hopping electrons are given by
\begin{equation}
          \dot{c_{i\sigma}} = - i/\hbar [ c_{i\sigma}, H ].
\end{equation}
For details of the analysis see Garcia {\it et al.} \cite{22}. Results are shown in Fig.13. These should be spin dependent for magnetic quantum dots,
$j_\sigma \propto \int d\varepsilon N_\sigma (\varepsilon)...$ .
Approximately, the DOS $N_\uparrow (\varepsilon)$ and
$N_\downarrow (\varepsilon)$ are split by the molecular field $H_{eff}$.

In the spirit of Onsager theory the field gradients drive
the tunnel current. Phases of the magnetizations are
also expected to play a role.
\begin{figure}
\centerline{\includegraphics[width=.55\textwidth]{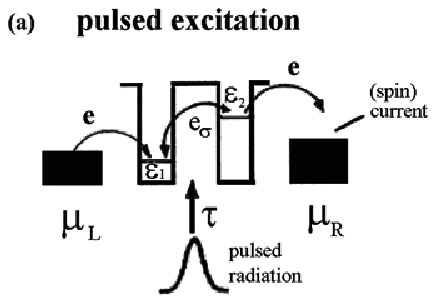}}
\centerline{\includegraphics[width=.55\textwidth]{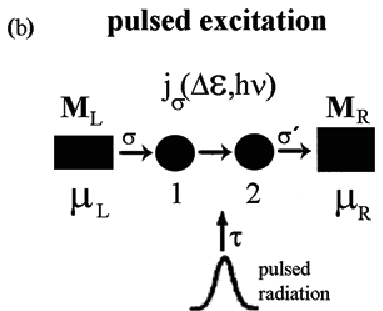}}
\caption{Illustration of a current between two spin polarized quantum dots (with electron states described for example by $\varepsilon_{i\sigma}=\varepsilon_i^0+U_{i}n_i\protect\overline{\sigma}$ +..., see Hubbard hamiltonian) coupled to two magnetic reservoirs (L,R). Optical manipulation of the occupations $n_{i\sigma}$ and of the current and its dynamics is of special interest and offers interesting physical behavior. The photon assisted tunneling is generally dependent on the energy barrier between the quantum dots, electron spin and light polarization and form of the pulsed radiation field. (a) Illustration of photon assisted quasi single electron hopping between quantum dots and (b) Sketch of many electron photon
assisted (spin dependent) tunnel current between two magnetic quantum dots (clusters)}
\end{figure}

In particular for tunneling involving not many electrons the currents $j=j_\uparrow + j_\downarrow$ and $j_s = j_\uparrow - j_\downarrow$ may exhibit
strong Rabi (v.St\"{u}ckelberg) oscillations due to back and
forth motion of the electrons. Results by M.Garcia {\it et al.}
are shown in Fig. 13 \cite{22}. Note, one estimates that approximately the currents $j_\uparrow$ and $j_\downarrow$ are split proportional to the field $H_{eff}$.
\begin{figure}
\centerline{\includegraphics[width=.55\textwidth]{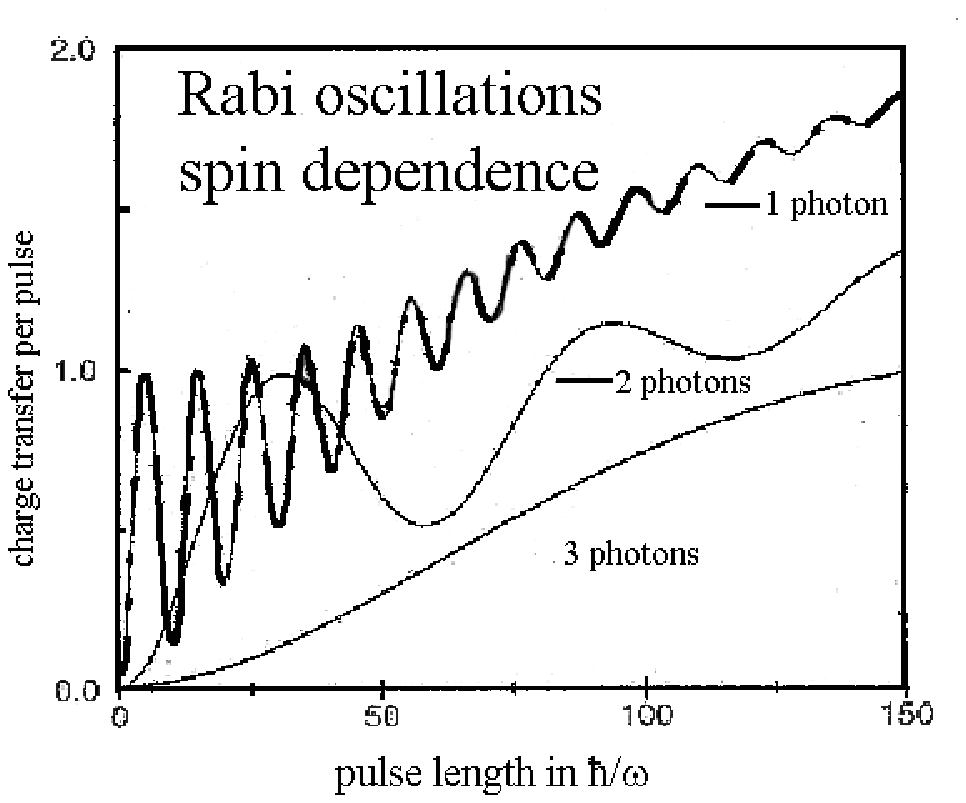}}
\caption{Spin dependent tunnel currents between two magnetic quantum dots, note the v. St\"{u}ckelberg (Rabi) oscillations. The
molecular field $H_{eff}$ splits spin up and spin down results. Of interest is the dependence of the oscillations on the duration of the photon field which controls the photon absorption during time of tunneling.}
\end{figure}

Note, the v. St\"{u}ckelberg (Rabi)-- oscillations with frequency approximately given by $\Omega \simeq 2\omega J_N (\frac{a}{\hbar\omega})$, with $N\hbar\omega = \sqrt{\Delta \varepsilon^2 + 4\omega^2}$, depend for a general field not periodic in time also on the shape of the potential of the photon field (and not only on its amplitude). $J_N$ is the Bessel function and N refers to the number of photons absorbed during hopping between quantum dots in order to fulfill above resonance condition \cite{22}. Note, the oscillations are large for short pulse duration and get damped for long pulses.

Apparently, the interplay of tunnel time and pulse time yields interesting behavior and controls the dynamics. The charge (or spin) transferred between the magnetic quantum dots depends on the number of photons absorbed during tunneling. For long pulse times this transfer increases when fewer photons are absorbed.

Of course, all thermoelectric and thermomagnetic effects
occur also for such a system of quantum dots. For two magnetic quantum dots, see Fig.12, the gradients ($T_1 - T_2$), ($\varepsilon_1 - \varepsilon_2$), or light field gradient might drive interesting currents. The fields generated by $j_{s} (t)$ and charge transfer are given by the Maxwell equations. The magnetic configuration of the quantum dots, parallel magnetizations or antiparallel magnetizations, is likely important
for spin or charge transfer (see GMR or TMR ).
In particular for intense photon fields one expects also that the polarization of the photons gets important.

The results can also be related to tunneling in molecules involving a few electronic states determining the tunneling and tunneling between molecules and surface of a (magnetic) solid and tunneling between two molecules or atomic clusters.

\subsubsection{Lattice of Quantum Dots}
First Anti--Quantum Dot Lattice: Currents in a system of magnetic quantum dots, for example ensemble of anti--quantum dots arranged as lattice, see Fig.12 for illustration, may exhibit interesting behavior. Applying the extension of Balian--Bloch theory by Stampfli {\it et al.} \cite{16} one gets for the DOS of the electrons
scattered by the anti--quantum dots the expression
\begin{eqnarray}
\Delta N(\varepsilon, B) &=& \sqrt{2}(a-d)/k_1 \pi \sum_{t=1}^\infty (\sinh \varphi/\sinh (4t+1)\varphi)^{1/2}\cos(BS) \nonumber \\ &\times& \sin\phi_{t,p}\exp-4tk_2(k_1/B)\varphi_1  + ..... \quad ,
\end{eqnarray}
with ($R_c = k/B$, which is the radius of the corresponding cyclotron radius), $\cosh \varphi = \frac{2a}{d} - 1$, and phase
\begin{equation}
   \phi_{t,p} = [ k_1(k_1/B)\varphi_1 + B/2(k_1/B)^2 (\varphi_1-\sin \varphi_1)+\delta_{t,4t} ] + \pi/2.
\end{equation}
Here, $\varphi_1 = 2 \arcsin \frac{a-d}{\sqrt{2}(k_1/B)}, S = 2t(a-d)^2$ and phase $\delta_{t,p}$ describing potential scattering at the surface of the quantum dot, $\delta_{t,p}= -\pi, for U \rightarrow - \infty$, see Stampfli {\it et al} \cite{16}.

Roughly, the states of spin up and spin down electrons are split by $H_{eff}$ and thus the DOS of spin up and spin down electrons are shifted by
the molecular field $H_{eff}$. If both B and $H_{eff}$ are perpendicular to the lattice of quantum--dots, then the molecular field may act similarly as the external magnetic field B.
\begin{figure}
\centerline{\includegraphics[width=.5\textwidth]{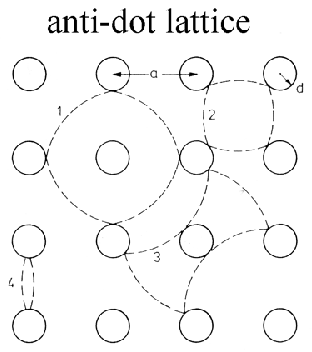}}
\caption{Polygonal paths 1, 2, 3 etc. in an anti-dot lattice resulting from spin dependent electron scattering by the repulsive potential of anti-dots (open circles)}
\end{figure}
The factor $\cos(SB)$ causes periodic oscillations in the electronic DOS, and for small B it is $R_c\propto 1/B$. For increasing B one has Landau--level
oscillations with periodicity proportional to (1/B), since $S\propto R_{c}^2$ and flux $\phi_2 \sim SB$ and $\Delta N(\varepsilon, \sigma, B)\propto\cos(SB)$. The DOS changes at $T_c$, which is the (Curie) ordering temperature of the ensemble of magnetic quantum dots. In strong magnetic fields B
the spin splitting of the electronic levels not only due to $H_{eff}$, but also due to B must be taken into account.

Note, it follows approximately for the electron current using $j\propto dF/d\phi$ that
\begin{equation}
j \sim \sin(BS).
\end{equation}

In Fig. 15 typical results are shown for the electronic structure of electrons scattered by an ensemble of anti--quantum dots. Such structure is reflected,
for example, by the magnetoresistance $\rho_{xx}$ and $\rho_{xx}\propto f[ N(\varepsilon, \sigma, B, ..)^2]$.
\begin{figure}
\centerline{\includegraphics[width=.6\textwidth]{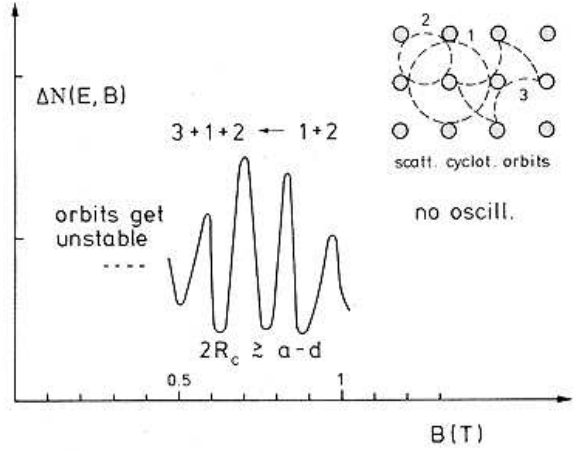}}
\caption{Schematic illustration of electronic DOS oscillations as a function of magnetic field B of electrons scattered by repulsive (square--well) potentials of quantum--dots. The oscillations depend on parameters a, d, see Fig.14, and molecular field $H_{eff}$ in case of magnetic anti--quantum dots, and magnetic ordering of the anti--quantum dots. Approximately, the DOS are spin split by $H_{eff}$.}
\end{figure}

A potential gradient or thermal gradient etc may cause a spin dependent flow, electron current in the anti--quantum dot lattice. Such currents can be described by Onsager theory and may be assisted by a photon field.

Secondly Lattice of Quantum Dots: In a lattice of quantum dots, see Fig.14 for illustration, electron hopping as described by the Tight--Binding hamiltonian or tunneling may occur between the quantum dots. Again, this may be assisted by a photon field. Then, a lattice of quantum dots between two metallic reservoirs may act as a switch due to varying the conductivity of the quantum dot lattice with the help of the photon field. Thus ultrafast dynamics may occur.

Force gradients $X_{i\sigma}$ induce
spin dependent currents in a lattice of magnetic quantum dots. Thermoelectric and thermogalvanic effects are expected. In particular the currents between two ( or many ) quantum dots having different magnetization, potential, temperature etc. may display interesting behavior.

\section{Summary}

Various experiments can be used to determine the generally spin dependent Onsager coefficients $L^\sigma$. The spin dependent forces $X^\sigma$
can generally be manipulated by light creating hot electrons and thus changing the various gradients $\Delta T$, $\Delta M$, etc.. For the tunnel system shown in Fig.2 one might expect interesting behavior if for example the metal 3 is replaced by Ce which electronic properties, valency changes upon photon induced and controlled population of the $s,d$ and $f$--states. Similarly currents change dramatically if in Fig.2 the material 3 consists of semiconductors like Si, Ge etc or magnetic semiconductors which conductivity is strongly affected by hot electrons.

Note, the Onsager
coefficients are given by the current current correlation function \cite{9,14,15}
\begin{equation}
          L_{il}(t) = \langle j_{i}(t) j_{l}(0) \rangle \quad ,
\end{equation}
with $j_{i}(t)$ calculated using response theory, Heisenberg or v.Neumann equation of motion ($\dot{\rho} \propto [\rho,H]+...$) or as a functional from the free--energy $F$ \cite{9,14,15}, see also Bloch using $j V = dF/dt$ with $V$ being the potential associated with the force $X_{i\sigma}(t)$. Then, (see response theory)

\begin{equation}
          j_{i}(t) = \int_\infty^t dt{'}L_{il}(t - t^{'}) X_{l}(t{'}),
\end{equation}
where the $L_{il}(t)$ are now calculated within an electronic theory. The analysis is simplified if $L_{ij} \propto \delta (t-t^{'})$ (Markov processes, see Kubo \cite{9}).

An electronic theory may be useful in order to apply field theoretical arguments and symmetry considerations, see for example Nogueira \cite{14}. The currents are
directly also calculated as derivatives of the electronic free--energy \cite{8}. One expects generally the formulae  $j V_{th} = dF/dt$, where $V_{th}$ denotes the "potential" associated with the driving force $X_{i\sigma(t)}$ of thermodynamics \cite{9}. Note, for large gradients $X_\sigma$ nonlinear contributions to the
currents may play a role.

In case of magnetic multilayer structures, for example a ferromagnet A on a ferromagnet B, one may induce spin currents by shining light on the surface of the thin magnetic film. This creates hot electrons and a temperature gradient and thus induces a spin current etc.. Note, according to Maxwell equations
$\Delta \overrightarrow{M(t)}$ will generate electric fields $E_i$, $i=x,y,z$, which should be reflected in the observed currents.

Onsager theory applies also to currents in (gases) liquids of magnetic ions and in the presence of a polarizing external magnetic field. Separating for example, such a system by a wall with appropriate holes into two compartments A and B, one may induce charge and spin currents driven by $X_{i\sigma}$,
voltage gradient, thermal gradients, magnetic field gradient, etc. . ( Spin--size effects may be a novel phenomenon. ) One gets approximately for the currents through the wall holes, see for example Kubo \cite{9},
\begin{equation}
         j_{\sigma}^\alpha \propto p_{\sigma}^\alpha /\sqrt{T^\alpha} + ... ,   \alpha = A, B,
\end{equation}
where $p_\sigma$ is the partial pressure due to ions with spin $\sigma$. Obviously, while for the stationary state $j_{i}^A = j_{i}^B$, gradients
$X_{i\sigma}$ cause corresponding currents. For example, in case of no pressure difference charge ( and spin ) may flow from A to B due to a temperature gradient and $T_A < T_B$.

Onsager theory has many applications in thermodynamics. As an example note, in case of magnetostriction thermodynamics yields \cite{9}
\begin{equation}
         \Delta M = - \left.\frac{\partial V}{\partial H_{eff}}\right)_{p,T} \Delta p + ...
\end{equation}
Hence, a pressure gradient changes the magnetization and may cause spin current ($\Delta p \rightarrow \Delta M \rightarrow j_s$ ). In particular a time dependent pressure p(t) drives a magnetization dynamics.

Onsager theory could also be applied to spin currents in topological insulators and at the interface of semiconducors in the presence of strong magnetic fields. Interplay of spin--orbit coupling and magnetic field and magnetism of substrate of semiconductor should yield interesting results.

Regarding (magnetic) atoms on lattices, including tunnel junctions (structures), spin dependent gradient forces $X_i$ acting on the atoms may cause
currents and novel behavior. Photon assistance of atom or molecule tunneling might be of particular importance.

Coupled currents under the influence of a magnetic field and radiation fields play likely also an important role in interstellar and galactic interactions and could be treated using same theory as above, using Onsager theory for magnetic, ionic gases. Coupling of magnetic currents
to black holes might yield interesting behavior \cite{23}.

Finally, interesting and novel behavior may occur if ( diffusion ) currents are accompanied by coupled chemical reactions, see de Groot \cite{9}. Then in magnetic systems spin and magnetization may play a role and cause magnetic effects \cite{23}.

This discussion demonstrates the many options for inducing spin currents and the powerful general analysis Onsager theory offers.

\section{Acknowledgement}

I thank C. Bennemann for help and many useful and critical discussions. This study is dedicated to Prof.J.B. Ketterson (USA) and Prof.V. Bortolani (Italy) for lifelong help, suggestions and general assistance. Last but not least I thank in particular F. Nogueira and M.Garcia for ideas, results and interesting discussions.


\begin{thebibliography}{99}
\bibitem{11} Note, taking into account magnitude and phase of the spin magnetization one gets generally $j_M \propto \Delta |M| +
             b_1 |M|\nabla \exp i\psi$ and similarly for $j_\sigma$ Josephson like currents. Of course, spin flip scattering
             plays a role and should be considered.
\bibitem{20} Note, the Landau--Lifshitz equation, $\frac{d\overrightarrow{M}}{dt} =  \gamma \overrightarrow{M}\times \overrightarrow{H_{eff}}+\alpha \overrightarrow{M}\times \frac{d\overrightarrow{M}}{dt} + \beta \overrightarrow{j}\times \partial_t\overrightarrow{M}$, includes damping. Approximately, damping is given by  $j_s \propto f(x/l_s) \propto \exp - (x/l_s)$, where $l_s$ denotes the spin mean free--path.
    A time dependent spin current yields in accordance with the Maxwell equations an electric field.
\bibitem{23} To include chemical reactions occurring in (open) magnetic systems with diffusion currents etc. one may extend Onsager theory taking into account spin and magnetization. For simplicity one may first neglect nonlinear behavior. (However, note frequently this may not be valid.). Then, $j_i=L_{ij}X_j+...$ and chemical reactions are taken into account by the forces $X_j=A_{j}/T$ with spin dependent chemical affinities $A_j=-\sum_{k} \nu_{k,j}\mu_k$. Here, the substance k with chemical potential $\mu_k$ is also characterized by its spin. The stoichometrical coefficients $\nu_{k,j}$ characterize the chemical reaction which couples to the currents. It is straightforward to work out details of the theory, see de Groot, Landau. Thus one gets spin dependent coupling of diffusion currents and chemical reactions, for example effects due to concentration gradients depending on spin and magnetization, etc.. Regarding intra- and intergalactic magnetic currents, it might be that black
    holes act on the currents similarly as chemical reactions on diffusion currents.
\end{thebibliography}
\end{document}